\documentclass[final,5p,times,twocolumn]{elsarticle}

\usepackage{graphics}
\usepackage{amsmath}
\usepackage{textcomp}

\journal{Nuclear Physics B}

\begin{document}

\begin{frontmatter}

\title{Asymmetry Dependence of the Nuclear Caloric Curve}

% WAITING FOR COMMENTS: Aldo, Kris, Zach, Paola

\author[AddrTron]{A.B. McIntosh} \ead{amcintosh@comp.tamu.edu}
\author[AddrTron,AddrLNS]{A. Bonasera}
\author[AddrTron,AddrChem]{P. Cammarata}
\author[AddrTron]{K. Hagel}
\author[AddrTron,AddrChem]{L. Heilborn}
\author[AddrTron,AddrChem]{Z. Kohley\fnref{AddrMSU}}
\author[AddrTron]{J. Mabiala}
\author[AddrTron,AddrChem]{L.W. May}
\author[AddrTron]{P. Marini\fnref{AddrGANIL}}
\author[AddrTron,AddrChem]{A. Raphelt}
\author[AddrTron,AddrAthens]{G.A. Souliotis}
\author[AddrTron,AddrChem]{S. Wuenschel\fnref{AddrWC}}
\author[AddrTron,AddrChem]{A. Zarrella}
\author[AddrTron,AddrChem]{S.J. Yennello}

\address[AddrTron]{Cyclotron Institute, Texas A$\&$M University, College Station, Texas, 77843, USA}
\address[AddrLNS]{Laboratori Nazionali del Sud, INFN, I-95123 Catania, Italy}
\address[AddrChem]{Chemistry Department, Texas A$\&$M University, College Station, Texas, 77843, USA}
\address[AddrAthens]{Laboratory of Physical Chemistry, Department of Chemistry, National and Kapodistrian University of Athens, Athens GR-15771, Greece}

\fntext[AddrMSU]{Present Address: National Superconducting Cyclotron Laboratory, Michigan State University, East Lansing, Michigan, 48824, USA}
\fntext[AddrGANIL]{Present Address: GANIL, Bd H Becquerel, BP 55027-14076, Caen, France}
\fntext[AddrWC]{Present Address: Weatherford College, Weatherford, TX, 76086, USA}

\begin{abstract}
A basic feature of the nuclear equation of state
is not yet understood:
the dependence of the nuclear caloric curve on the neutron-proton asymmetry.
Predictions of theoretical models differ on the magnitude
and even the sign of this dependence.
In this work,
the nuclear caloric curve is examined for
fully reconstructed quasi-projectiles around mass $A=50$.
The caloric curve extracted
with the momentum quadrupole fluctuation thermometer
shows that the temperature varies linearly with
quasi-projectile asymmetry $\frac{N-Z}{A}$.
An increase in asymmetry of 0.15 units
corresponds to a decrease in temperature on the order of 1 MeV.
These results also highlight the importance of a full quasi-projectile reconstruction
in the study of thermodynamic properties of hot nuclei.
\end{abstract}


\end{frontmatter}

\section{Introduction}

The relation between the temperature and the excitation energy of a system
(the caloric curve)
is of fundamental importance in a wide variety of physical systems.
Since the application of the concept of a caloric curve to
atomic nuclei \cite{Bethe37,Weisskopf37},
several ``thermometers'' have been used
to elucidate properties of excited nuclei
including the transition from evaporative-type decay to nuclear multifragmentation
(see \cite{Shlomo05,Kelic06}, and references therein).
Recently,
a clear mass-dependence of the caloric curve for finite nuclei
has been demonstrated \cite{Natowitz02}.

The dependence of the nuclear caloric curve
on the neutron/proton asymmetry, $\frac{N-Z}{A}$,
remains uncertain due to conflicting predictions from theoretical calculations
and the relatively small body of experimental data on the subject.
Some theoretical approaches
predict that critical temperatures or limiting temperatures
would be higher for neutron-poor systems \cite{Kolomietz01,Hoel07};
others
predict higher temperatures for neutron-rich systems \cite{Besprosvany89,Ogul02,Su11}.
Inclusion (or intentional omission) of a ``gas'' phase
that interacts with the bulk system
is expected to impact the asymmetry dependence
of the temperature of the bulk system \cite{Hoel07,Besprosvany89}.
The observation of an asymmetry dependence
may support the physical picture of a nuclear liquid interacting with its vapor \cite{Hoel07},
or may allow insight into the driving force of nuclear disassembly \cite{Sfienti09}.
Studies in recent years \cite{Colonna06,Shetty07,Marini12}
have sought to probe the asymmetry energy
in the nuclear equation of state
by examining the fragments produced in heavy-ion reactions.
Since these studies often assume the temperature
is independent of the asymmetry,
observation and characterization of an asymmetry-dependence of the caloric curve
would allow a refined interpretation of fragment yield data
(e.g. in the statistical interpretation of isoscaling).
Moreover, characterization of this asymmetry dependence may offer the opportunity
to probe the asymmetry energy in a new way;
this is discussed below.
Experimentally, temperatures
have shown either a small dependence \cite{Sfienti09,Trautmann08}
or no discernible dependence \cite{Wuenschel10,Wuenschel09_Thesis}
on the asymmetry of the initial system.

Motivation for an asymmetry dependence of the nuclear temperature
may be seen in the following argument
based on Landau theory \cite{Huang87,Bonasera08,Huang10}.
We consider a fragmenting nuclear source,
and write the free energy per nucleon of each fragment produced by the source as
\begin{equation}
 \left( \frac{F}{A} \right) _{f}
 = \left( \frac{F}{A} \right) _{f,0} + H m_f + V_c Z_s Z_f + \frac{3}{2} T
\end{equation}
where $\left( \frac{F}{A} \right) _{f,0}$ is the free energy per nucleon of the fragment in isolation
and $T$ is the temperature of the system.
The asymmetry of the fragment $m_f = \frac{N_f-Z_f}{A_f}$ increases the free energy
in proportion to $H$, the conjugate variable of $m_f$.
The quantity $H = c_{asy} m_s$
is the asymmetry field due to the source
where
$c_{asy}$ is the asymmetry energy coefficient
and
$m_s = \frac{N_s-Z_s}{A_s}$ is the asymmetry of the source \cite{Bonasera08}.
The Coulomb interaction between 
the charged fragment of interest ($Z_{f}$) and the remainder of the source ($Z_{s}$)
is described by the $V_{c}$ term.
Consider two identical fragments produced from
two sources with the same mass and excitation energy but different asymmetry.
Taking the difference in the free energy
for these two fragments,
$\left( \frac{F}{A} \right) _{f,0}$ cancels exactly.
This gives a linear dependence of temperature on the source asymmetry:
\begin{equation}
 \label{eq:DeltaT}
 \Delta T = \Delta m_s \left( \frac{1}{3} V_{c} A_{s} Z_{f} - \frac{2}{3} m_f c_{asy} \right)
            + \frac{2}{3} \Delta \left( \frac{F}{A} \right)_{f}.
\end{equation}
In the present work,
we demonstrate such an asymmetry dependence
of nuclear temperatures exists.

\section{Experiment and Event Selection}

To investigate the dependence of the nuclear caloric curve on asymmetry,
heavy-ion collisions at intermediate energy were studied.
Charged particles and free neutrons produced in reactions of
$^{70}$Zn+$^{70}$Zn,
$^{64}$Zn+$^{64}$Zn, and
$^{64}$Ni+$^{64}$Ni
at E/A = 35 MeV \cite{Kohley10,Kohley11_LCP}
were measured with excellent isotopic resolution in the
NIMROD-ISiS $4\pi$ detector array \cite{Wuenschel09_Thesis,Wuenschel09_NIM}.
The quasi-projectile
(QP, the primary excited fragment
that exists momentarily after a non-central collision)
was reconstructed,
including determination of the QP \emph{composition} (both A and Z).
Excitation energies above $E^*/A$ = 2 MeV
are well measured with this setup.

\begin{figure}
\includegraphics{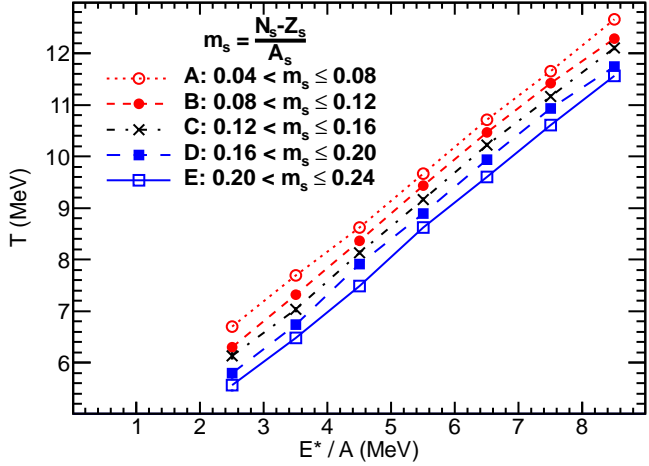} 
\caption{\label{fig:TFlucProt}
(Color online)
Caloric curves for isotopically reconstructed sources
with mass $48 \leq A \leq 52$,
extracted with the momentum quadrupole fluctuation method.
Each curve corresponds to a narrow range in source asymmetry, $m_s$.
}
\end{figure}

The uncertainty in the composition arises mainly from the free neutron measurement,
which arises from the efficiency of the neutron detector (70\%)
and from background signals in the neutron detector.
The background signals were measured;
the variance of the background multiplicity is a factor of 10 smaller than the
variance of the raw neutron multiplicity.
The excitation energy was deduced using
the measured free neutron multiplicity,
the charged particle kinetic energies,
and the Q-value of the breakup.
Use of simulations of the detector response \cite{Marini12b} indicate
that the uncertainty in the excitation energy per nucleon due to the uncertainty in the free neutron multiplicity
is around 0.1 MeV, which is significantly smaller than the spacing between even the closest caloric curves.
This uncertainty does not bias the results presented in this letter.

Building on previous work \cite{Marini12,Wuenschel10,Wuenschel09_Thesis,Wuenschel09_Isoscaling},
three cuts are made to select equilibrated QP sources.
To exclude fragments that clearly do not originate from an equilibrated QP source,
the fragment velocity in the beam direction $v_z$,
relative to the velocity of the heavy residue $v_{z,PLF}$,
is restricted.
The accepted window on $\frac{v_z}{v_{z,PLF}}$ is $1 \pm 0.65$ for Z=1,
$1 \pm 0.60$ for Z=2,
and $1 \pm 0.45$ for Z$\geq$3.
The mass of the reconstructed QP is required to be $48 \leq A \leq 52$.
To select QPs that are equilibrated,
it is required that the QP be on average spherical.
This is achieved with a selection on the longitudinal momenta $p_z$
and transverse momenta $p_t$
of the fragments comprising the QP:
$-0.3 \leq log_{10}(Q_{shape}) \leq 0.3$
where $Q_{shape} = \frac{\sum{p_z^2}}{\sum{p_t^2}}$
with the sums extending over all fragments of the QP.
Since the shape degree of freedom is slow to equilibrate,
these QPs that are on-average spherical should be thermally equilibrated.
Over the range of excitation energies presented in this work,
the typical QP is comprised of one large fragment, several light particles (Z$\leq$2) and one IMF (3$\leq$Z$\leq$8).

The temperatures of the QPs are calculated
with the momentum quadrupole fluctuation method \cite{Zheng11_PLB},
which has been previously used to examine
temperatures of nuclei \cite{Wuenschel10,Wuenschel09_Thesis,Stein12X}.
The momentum quadrupole is defined as
$Q_{xy} = p_{x}^{2} - p_{y}^{2}$
using the transverse components $p_x$ and $p_y$ of the particle's momentum
in the frame of the QP source.
Assuming a Maxwell-Boltzmann distribution,
the variance of $Q_{xy}$ is related to the temperature by
$\langle \sigma_{xy}^2 \rangle = 4m^2T^2$
where $m$ is the probe particle mass \cite{Wuenschel10, Zheng11_PLB}.
For this analysis,
protons, which are abundantly produced in the collisions,
are used as the probe.
The longitudinal component, $p_{z}$, is excluded
to minimize any contribution from
the collision dynamics, which manifests in the beam direction.
The effects of secondary decay
on this thermometer should be small \cite{Wuenschel10}:
the thermal energy in the primary clusters
is significantly less than that in the QP,
so the width of the momentum quadrupole
is dominated by the QP breakup.

\begin{figure}
\includegraphics{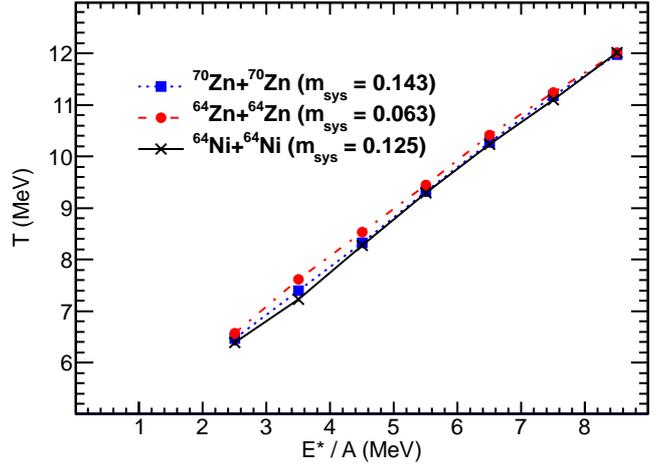} 
\caption{\label{fig:CompareSystems_Prot_OnePanel}
(Color online)
Caloric curves for isotopically reconstructed sources
with mass $48 \leq A \leq 52$,
extracted with the momentum quadrupole fluctuation method.
Each curve corresponds to a selection on the asymmetry of the initial system.
}
\end{figure}

\section{Results and Discussion}

Figure \ref{fig:TFlucProt}
shows the temperature as a function of excitation energy per nucleon ($E^*/A$)
of the QP as determined with the momentum quadrupole fluctuation thermometer
using protons as the probe particle.
Data points are plotted for 1MeV-wide bins in excitation energy per nucleon.
For clarity, the points are connected with lines to guide the eye.
The error bars correspond to the statistical uncertainty
and where not visible are smaller than the points.
The temperature shows a monotonic increase with excitation energy.
At $E^*/A$ = 2.5 MeV,
the temperatures are around 6 MeV;
by $E^*/A$ = 8.5 MeV,
the temperatures have risen to around 12 MeV.
Each curve corresponds to a narrow selection in
the asymmetry of the source, $m_s = \frac{N_s-Z_s}{A_s}$,
as indicated in the legend.
The average asymmetries for the selections are
0.0640, 0.0988, 0.1370, 0.1758, and 0.2145.
The caloric curve is observed to depend on
the asymmetry of the source.
Increasing the neutron content of the QP source
shifts the caloric curve to lower temperatures.
In fact, the caloric curves for the five source asymmetries
appear parallel and equally spaced.
An increase of 0.15 units in $m_s$
corresponds to a decrease in the temperature
by about 1.1 MeV.

In previous data,
a plateau in the caloric curve has been observed
and interpreted as a signature of a phase transition \cite{Kelic06,Natowitz02}.
In the present data, there is no plateau observed
in the caloric curve for these excited sources.
However, this is not unexpected
for such small sources ($A \approx 50$)
where the plateau is not as well defined
as it is for heavier sources \cite{Natowitz02},
and may be masked by a varying density \cite{Borderie12,Mabiala12}.
Moreover,
the plateau might lie entirely above
the excitation energies measured in this experiment \cite{Natowitz02}.

\begin{figure}
\includegraphics{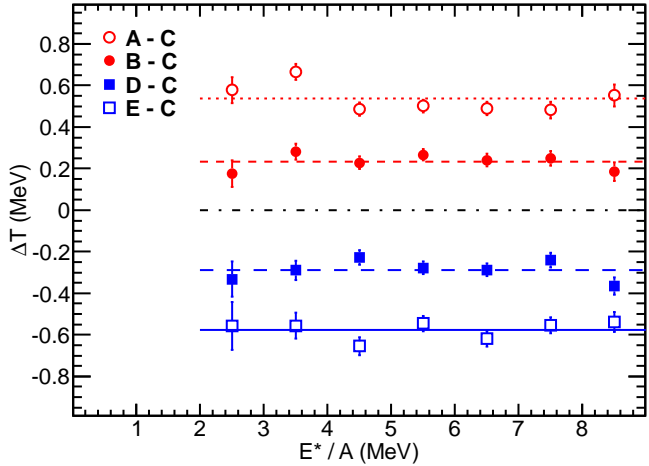} 
\caption{\label{fig:TFlucProtDiff}
(Color online)
Temperature difference of each caloric curve
(relative to the middle caloric curve, $0.12 < m_s \leq 0.16$).
The horizontal lines correspond to the average difference over the indicated range in excitation energy.
The labels ``A-C'', etc., refer to the $m_s$ bins in Fig.~\ref{fig:TFlucProt}.
}
\end{figure}

The composition of the de-exciting QP changes with time.
Ideally, the probe particles should be correlated with the composition of the source
at the time of their emission.
Our method determines the initial composition of the QP,
and correlates this with the temperature deduced using the measured particles
which are emitted over a range of times.
Correlating the initial composition of the QP, rather than the instantaneous composition,
could mask the asymmetry dependence of the temperature to some extent.
Thus, the true dependence may be even greater than
the considerable asymmetry dependence of nuclear temperatures observed here.

The asymmetry of the decaying source
should impact the caloric curve more directly
than the asymmetry of the initial system.
We have explored the impact of selecting on the
asymmetry of the initial system
rather than the asymmetry of the reconstructed source.
Figure \ref{fig:CompareSystems_Prot_OnePanel}
shows the proton momentum quadrupole fluctuation caloric curves
selected on the asymmetry of the initial system.
Data points are plotted for each 1MeV-wide bin in excitation energy.
The error bars correspond to the statistical uncertainty,
and where not visible are smaller than the points.
The temperatures for the neutron-rich systems are generally the lowest.
This is consistent with the results presented above
(Fig.~\ref{fig:TFlucProt}),
though the magnitude of the effect is much smaller.
Here, decreasing the \emph{system} asymmetry from 0.143 to 0.063
increases the temperature by about 0.1 MeV,
while decreasing the \emph{source} asymmetry by this amount
increases the temperature by about 0.55 MeV.
Though the asymmetry of the initial system might be employed as a surrogate
for the asymmetry of the fragmenting system,
at intermediate beam energies
these are only weakly correlated.
The asymmetry distribution
of the QP following the interaction
is broad ($\sigma_{m_s} \approx 0.07$);
there is considerable overlap in the distributions
from the different initial systems,
in agreement with previous work \cite{Marini12,Wuenschel09_Thesis,Wuenschel09_Isoscaling,Rowland00,Rowland03,Galanopoulos10}.
The present data show that
the ability to select the composition of the QP source
provides much greater sensitivity to the asymmetry dependence of the caloric curve.

The shift in temperature due to the asymmetry of the source
is examined more closely in Fig.~\ref{fig:TFlucProtDiff}.
The central curve from Fig.~\ref{fig:TFlucProt} ($0.12 < m_s \leq 0.16$) has been used as a reference.
The difference in temperature between each caloric curve and the reference
is plotted as a function of the excitation energy.
Error bars corresponding to statistical uncertainties are shown,
and where not visible are smaller than the points.
The differences in the momentum quadrupole fluctuation temperature
are fairly constant.
The shift in the caloric curve
with asymmetry is essentially independent of excitation energy
over the range measured here.
The average $\Delta T$ for each pair of $m_s$ bins is indicated by the horizontal lines.
Figure~\ref{fig:TFlucProtDiff} shows even more clearly than Fig.~\ref{fig:TFlucProt}
that the caloric curves for different $m_s$ bins are parallel and equally spaced.

\begin{figure}
\includegraphics{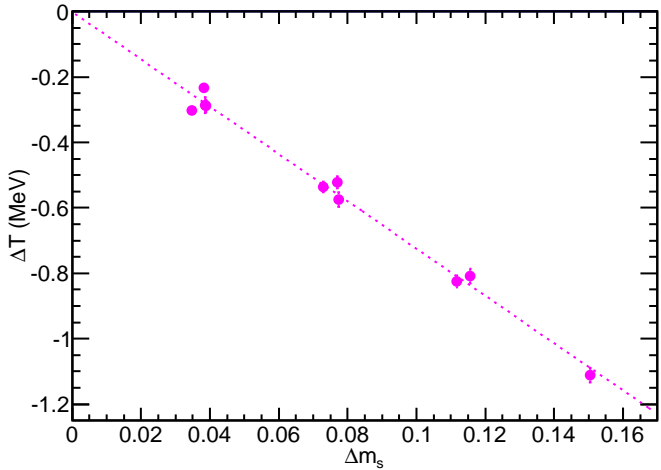} 
\caption{\label{fig:TempDiff_DeltaTvsMs_JustProt}
(Color online)
Change in temperature as a function of the source asymmetry $m_s = \frac{N_s-Z_s}{A_s}$.
The dashed line is a linear fit to the experimental data.
}
\end{figure}

The temperature shift $\Delta T$ of the caloric curve
is plotted as a function of changing source asymmetry $\Delta m_s$
in Fig.~\ref{fig:TempDiff_DeltaTvsMs_JustProt}.
This is obtained in the following way.
For each possible pairing of the five caloric curves (10 pairings total),
the temperature difference as a function of excitation energy is calculated.
These differences show no trend with excitation energy,
(as evidenced in Fig.~\ref{fig:TFlucProtDiff});
the average of the temperature difference is taken as $\Delta T$.
For each possible pairing of the five caloric curves,
$\Delta m_s$ corresponds to the difference in the mean asymmetries
of the pair of curves.
The error bars show the statistical errors,
and where not visible are smaller than the points.
A negative correlation of $\Delta T$ with $\Delta m_s$ is observed,
and is well described by a linear fit
over the broad range in source asymmetry,
with slope -7.3 MeV.

This correlation is consistent with the free energy argument presented (Eq.~\ref{eq:DeltaT}).
By averaging over excitation energy,
we isolate the dependence on $m_s$ from $\Delta\left(\frac{F}{A}\right)_f$;
the intercept is zero here by construction.
The slope of the correlation is dependent on the asymmetry and Coulomb terms.
It may be possible to extract
information on the asymmetry energy coefficient.
For this, an accurate estimate of the density would be necessary,
which will be the focus of future work.

The presently observed decrease in temperature with increasing asymmetry
can be compared to the recent result from the ALADIN collaboration \cite{Sfienti09,Trautmann08}
which shows for peripheral collisions
a modest dependence of the temperature
on the neutron-proton asymmetry.
In the ALADIN analysis, the asymmetry of the initial system is used.
The method used to select the composition
is important,
and is examined in the present work.
At intermediate energy,
the difference between the asymmetry of the source (Fig.~\ref{fig:TFlucProt})
and the initial system (Fig.~\ref{fig:CompareSystems_Prot_OnePanel})
shows the importance of the full source reconstruction, including the free neutrons:
selection of the source (rather than the initial system) allows greater sensitivity to the
asymmetry dependence of the temperature.
Though the evolution of this difference with increasing beam energy
up to E/A = 600 MeV of the ALADIN measurement remains to be seen,
observables dependent on the QP asymmetry
have shown a strong dependence with energy,
becoming more pronounced at lower excitation energy \cite{Veselsky01}.
Studies at various beam energies,
or studies sensitive to a broader range in excitation energy,
could probe a possible evolution with excitation energy
of the asymmetry dependence of the nuclear temperature.

\section{Conclusion}

In conclusion,
we have provided experimental evidence for a dependence
of nuclear temperatures on the neutron-proton asymmetry
using a kinetic thermometer.
Of crucial importance is the
selection on the asymmetry of the fully-reconstructed
isotopically-identified fragmenting source,
rather than the initial system asymmetry.
The temperature
is observed to depend
linearly on the source asymmetry.
The temperature changes on the order of 1 MeV
with varying asymmetry.
Future experimental studies should examine
particularly the low excitation energy region
to investigate how the asymmetry dependence changes
as the reaction mechanism evolves from
the onset of evaporation through multifragmentation;
and, over a wide range of excitation energy,
should constrain (or at least determine) the density
of the emitting source to allow a more quantitative understanding of the deduced caloric curves.

\section*{Acknowledgments}
We thank the staff
of the TAMU Cyclotron Institute for providing the
high quality beams which made this experiment possible.
This work was supported by
the Robert A. Welch Foundation (A-1266),
and the U. S. Department of Energy (DE-FG03-93ER-40773).

\bibliographystyle{model1-num-names}
\bibliography{AsyDepCal}

\end{document}